\documentclass[traditabstract]{aa}  

\usepackage{graphicx}
\usepackage{txfonts}
\usepackage{natbib}
\usepackage{ulem}

\def\kms{$\mathrm{km\, s^{-1}}$}

\begin{document} 

   \title{Chemical abundances of the metal-poor horizontal-branch stars \object{CS~22186-005} and \object{CS~30344-033}
   \thanks{Based on observations collected at the European Southern
   Observatory, Chile, Program IDs 077.D-0299 and 076.D-0546(A).}}
  
  \author{
  \c{S}. \c{C}al{\i}\c{s}kan\inst{1}
  \and E. Caffau\inst{2}\and P. Bonifacio\inst{2}\and N. Christlieb\inst{3} 
  \and L. Monaco\inst{4}\and T. C. Beers\inst{5,6}\and B. Albayrak\inst{1}
  \and L. Sbordone\inst{7,8,3}
}

\institute{
  Department of Astronomy and Space Sciences, Ankara University, 
  06100, Tando\u{g}an, Ankara, Turkey\\
  \email{seyma.caliskan@science.ankara.edu.tr}
  \and
  GEPI, Observatorie de Paris, CNR, Universit\'{e} Paris Diderot, Place Jules
  Janssen, 92190 Meudon, France\\ 
  \email{Elisabetta.Caffau/Piercarlo.Bonifacio@obspm.fr}
  \and
  Zentrum f\"{u}r Astronomie der Universit\"{a}t Heidelberg,
  Landessternwarte, K\"{o}nigstuhl 12, 69117 Heidelberg, Germany\\
  \email{n.christlieb@lsw.uni-heidelberg.de}           
  \and 
  European Southern Observatory, Casilla 19001, Santiago, Chile
  \email{lmonaco@eso.org}
  \and
  National Optical Astronomy Observatory, Tucson, AZ 85719, USA 
  \email{beers@noao.edu}
  \and
  JINA: Joint Institute for Nuclear
  Astrophysics, Michigan State University, East Lansing, MI~48824, USA
  \and
  Millennium Institute of Astrophysics, Av. Vicuna Mackenna 4860, 782-0436 Macul, Santiago, Chile
  \email{lsbordon@astro.puc.cl}
  \and
  Pontificia Universidad Católica de Chile, Av. Vicuna Mackenna 4860, 782-0436 Macul, Santiago, Chile\\
}
   \date{Received 19 June 2014/ Accepted 14 August 2014}

 
\abstract {We report on a chemical-abundance analysis of two very 
metal-poor horizontal-branch stars in the Milky Way halo: CS~22186-005 
($\mathrm{[Fe/H]=-2.70}$) and CS~30344-033 ($\mathrm{[Fe/H]=-2.90}$). 
The analysis is based on high-resolution spectra obtained at ESO, with the 
spectrographs HARPS at the 3.6\,m telescope, and UVES at the VLT. 
We adopted one-dimensional, plane-parallel model atmospheres assuming local thermodynamic equilibrium. 
We derived elemental abundances for 13 elements for CS~22186-005 and 
14 elements for CS~30344-033. This study is the first abundance analysis of 
CS~30344-033. CS~22186-005 has been analyzed previously, but we report 
here the first measurement of nickel (Ni; $Z = 28$) for this star, based on 
twenty-two \ion{Ni}{i} lines ($\mathrm{[Ni/Fe]}=-0.21\pm0.02$); the measurement is significantly
below the mean found for most metal-poor stars. Differences of up to 0.5\,dex in $\mathrm{[Ni/Fe]}$ ratios were determined by different
authors for the same type of stars in the literature, which means that it is not yet possible to conclude 
that there is a real intrinsic scatter in the $\mathrm{[Ni/Fe]}$ ratios. For the other elements 
for which we obtained estimates, the abundance patterns in these two stars match the Galactic trends defined by giant and turnoff stars well.
This confirms the value of horizontal-branch stars as tracers of the chemical properties 
of stellar populations in the Galaxy. Our radial velocities measurements for CS~22186-005 
differ from previously published measurements by more than the expected 
statistical errors. More measurements of the radial velocity of this star are 
encouraged to confirm or refute its radial velocity variability.}

  \maketitle
%
%

\section{Introduction}

In most instances (except in cases of, e.g., pollution from a binary
companion) the atmospheres of stars are expected to reflect the chemical
composition of the interstellar medium from which they were born, hence,
metal-poor stars provide "archaeological evidence" that constrains the
early epochs of star formation in our Galaxy\footnote{See, however,
\citet{hattorietal14}}. These stars thus play important roles in
understanding the nature of the first objects that formed in the
Universe; their chemical abundances can be used to infer the first few
steps of Galactic chemical evolution. 

Among the possible tracer populations of the very metal-poor
component(s) of the Galactic halo, horizontal-branch (HB) stars have
received much less attention than giant-branch (GB) or main-sequence
turnoff (MSTO) stars. However, in recent years a significant number of
HB stars have been subject to detailed chemical analyses
\citep{prestonetal06, forsneden10, For11, Hansen11}. These studies have
shown that except for Li and C (which can be readily altered during a
star's ascent of the red giant branch), these stars are indeed reliable
fossil records of the material from which they formed. The HB stars are
more luminous than the MSTO stars and can thus be used to probe the
Galaxy to larger distances. With respect to GB stars, HB stars have the 
clear advantage that their masses can be derived more accurately 
from their surface gravities and effective temperatures through comparison
with stellar evolutionary tracks.  

\begin{table*}[htbp]
   \caption[]{Properties and observations log of the two program stars.}
   \label{tab1}
   \centering
   \tiny
   \begin{tabular}{c c c c c c c c c c}
   \hline
   \noalign{\smallskip}
   Object & RA & DEC & B &Spectrograph & Setup &Obs. date & Exp. time & $\lambda$ range & Resolving power\\ 
   &[$\mathrm{h}$~$\mathrm{m}$~$\mathrm{s}$] & [$^\circ$~$^\prime$~$^{\prime\prime}$]& $\mathrm{[mag]}$ & & & & $\mathrm{[s]}$ & [{\AA}] & \\
   \noalign{\smallskip}   
   \hline
   \noalign{\smallskip} 
   CS~22186-005&$04~13~09.1$&$-35~50~38.7$&$13.33$&HARPS&       &March~10, 2006   &2x1800      &3778-6908&115000\\ 
               &            &             &       &UVES &blue390&November~18, 2005&3600        &3280-4560&71000 \\
               &            &             &       &     &       &                 &             &        &      \\               
               &            &             &       &     &blue390&May~28, 2006     &3600        &3300-4510&      \\ 
   CS~30344-033&$22~55~31.9$&$-34~42~59.0$&$14.61$&UVES &red580l&May~28, 2006     &3600        &4760-5800&40000 \\
               &            &             &       &     &red580u&May~28, 2006     &3600        &5820-6840&      \\ 
\noalign{\smallskip}
   \hline
   \end{tabular}                                   
   \end{table*}

In this paper we report on a detailed chemical-abundance analysis of two
very metal-poor HB stars, one of which is analyzed here for the first
time. CS~22186-005 was identified in the HK objective-prism survey of
\citet{beersetal92} and was classified as an HB star by \citet{norrisetal96}. 
The abundance analysis of this star was performed by \citet{prestonetal06} 
and \citet{forsneden10}. Broadband photometric data for CS~30344-033 
\citep[rediscovered in the Hamburg/ESO survey as HE~2252-3458;][]{christliebetal08} 
was published by \citet{norrisetal99}, who noted its status as a metal-weak candidate.
No abundance analysis has previously been performed for this star. Section
2 describes our spectroscopic observations and data reduction, while Section 3 includes comments about 
the possible radial velocity variation of CS~22186-005. Section 4 presents
details of our abundance analysis. In Section 5, we discuss the
assignment of the evolutionary stages of these stars and their
associated masses. Section 6 summarizes the results of our abundance
analysis, and Section 7 presents a discussion of these results. 
We conclude in Section 8.


\section{Spectroscopic observations and data reductions}
 
High-resolution spectra of CS~22186-005 were obtained with two different 
spectrographs: the HARPS 3.6\,m instrument and UVES at the VLT \citep{dekkeretal00}. 
Two HARPS spectra ($R\sim115,000$), covering a wavelength range of 3778 to 6908\,{\AA}, 
were acquired on March 10, 2006, each with an exposure time of 1800\,s. Both spectra were 
shifted to a rest-wavelength scale before they were combined. The spectrum of CS~22186-005, 
obtained with UVES, has a resolving power of $R\sim71,000$. It has a wavelength range of $3280-4560$\,{\AA}  and 
was acquired with 3600\,s on November 18, 2005. UVES at the VLT is used to obtain high-resolution spectra of CS~30344-033.  
The spectra have a spectral resolution of $R\sim40,000$. They were acquired with 3600\,s on May 28, 2006, 
covering a wavelength range of $3300-6840$\,{\AA}. We used the pipeline-reduced spectra (as provided
by the ESO data management division) retrieved from the ESO archive \footnote{http://archive.eso.org/wdb/wdb/eso/repro/form}$^{,}$
\footnote{http://archive.eso.org/wdb/wdb/adp/phase3\_spectral/\\form?phase3\_collection=UVES\_ECHELLE}. 
In Table~\ref{tab1} we list details of the observing sessions and instrumental setups as well as the
coordinates and brightness of the stars, observing dates, exposure
times, wavelength ranges of the spectra, and resolving powers.

\begin{figure}[htbp]
   \centering
   \includegraphics[bb=70 450 533 671,width=9cm,clip]{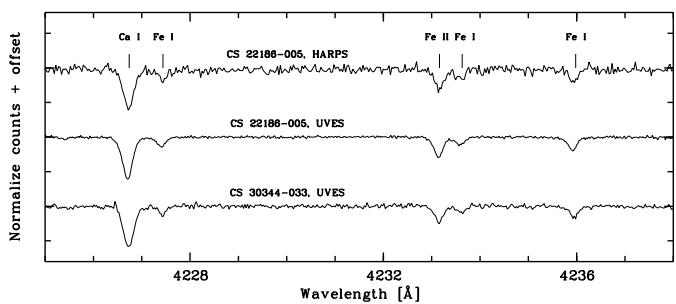}
\caption{The $4225-4238$\,{\AA} region of the final spectra of CS~22186-005 and CS~30344-033.}
         \label{spectraq}
   \end{figure}

As part of our reduction procedure, we rebinned the HARPS spectrum of CS~22186-005 
and the UVES spectrum of CS~30344-033 by a factor of two to increase their 
signal-to-noise ($S/N$) ratios. We did not apply any rebinning to the UVES spectrum of 
CS~22186-005, because its $S/N$ is sufficient. We list in Table~\ref{tab2} the $S/N$ ratios of each spectrum after rebinning 
in the wavelength regions around $4000$\,{\AA} and $5000$\,{\AA}.
In Fig.~\ref{spectraq}, we show a portion of the final spectra in the $4225-4238$\,{\AA} region for our two stars.
We measured the radial velocities of the stars from the \ion{Ca}{ii}~H\&K lines
(in the blue regions), the \ion{Mg}{i}~b lines (in the red regions), and
weak iron lines over the entire wavelength ranges, as $v= c(\lambda_{obs}-\lambda_0)/\lambda_0$, 
where $c$ is the speed of light and $\lambda_0$ is the rest wavelength of the line.
Then we averaged the radial velocities measured from the different lines. The resulting values
of the barycentric radial velocities and their errors are listed in
Table~\ref{tab2}. 
    
\section{Radial velocities: is CS~22186-005 a variable?}

For CS~30344-033, our radial velocity measurement  
reported in Table~\ref{tab2} is the only one available. 
For CS~22186-005, however, we have two spectra, and three other measurements
are available in the literature.
The HARPS and UVES spectra provide two measurements that are consistent
with no variation in the radial velocity, but only
four months separate the two observations.
From the low-resolution HK survey spectra, \citet{beersetal92} found 183$\pm 10$\,\kms
and \citet{beers2000} 192$\pm 10$\,\kms; given the lower precision, these measurements are consistent with ours.

\citet{prestonetal06}, reported 183\,\kms, but did not state an error. Considering
that their analysis is based on a high-quality MIKE spectrum, 
\citet{roedereretal14b} showed an error of $\pm 0.6-0.8$\,\kms for the radial velocity 
measurements of MIKE spectra, similar to our UVES spectrum.
At face value, this measurement is inconsistent with both our measurements
and would be evidence for a variation in the radial velocity of the 
star, unless there is a large offset in measured radial velocity between
MIKE and UVES, which is unlikely.

\begin{table*}[htbp]
   \caption{Measured $S/N$ ratios and barycentric radial velocities.}
   \label{tab2}
   \centering
   \tiny
   \begin{tabular}{c c c c c c}
   \hline
   \noalign{\smallskip}
   Object&Instrument&$S/N$&$S/N$ &$S/N$&$v_{\mathrm{bary}}$\\
         &&[@$4000$\,{\AA}]&[@$5000$\,{\AA}] &[@$6000$\,{\AA}] &$\mathrm{[km/s]}$\\
   \noalign{\smallskip}
   \hline
   \noalign{\smallskip}
   CS~22186-005&HARPS &$16$ &$30$    &$40$    &$193.6\pm1.1$\\
               &UVES  &$100$&$\cdots$&$\cdots$&$194\pm1$    \\
               &      &     &        &        &             \\
   CS~30344-033&UVES  &$40$ &$120$   &$90$    &$101.1\pm0.6$\\
   \noalign{\smallskip}
   \hline
   \end{tabular}
   \end{table*}

Additional monitoring of the radial velocity of this star is encouraged.
If the variations in radial velocity are confirmed, it might be a binary 
with a very long period (given the near coincidence of the UVES and HARPS 
radial velocities). This would make it a very interesting object, since 
we could obtain an estimate of its mass independent of evolutionary tracks. 
The star is bright and sufficiently nearby that the Gaia satellite could provide 
an astrometric orbit if the star were a binary.  
   
\section{Abundance analysis}
\subsection{Stellar parameters}

For each star we started from an initial estimate of the atmospheric parameters. 
For CS~22186-005, we used the parameters given by \citet{forsneden10}.
For CS~30344-033 we used the  colors given by \citet{christliebetal08} and the calibrations of
\citet{casagrandeetal10} for the effective temperature. The colors in 
\citet{christliebetal08} were derived from the objective prism spectra and are thus 
more uncertain than CCD photometry. From this temperature we used 
the Yale-Yonsei isochrones \citep{Y2} of 10 and 12\,Gyr and $\rm{[Fe/H]=-2.5}$, $\rm{[\alpha/Fe]= +0.3}$
to estimate the surface gravity. With these atmospheric parameters we computed a model atmosphere
using version~9 of the ATLAS code \citep{kurucz93a, kurucz05, sbordoneetal04} using the models
in the grid of \citet{castelli03} as starting model and the same opacity distribution functions.  
The computation assumed local thermodynamic equilibrium (LTE),
plane-parallel geometry, hydrostatic equilibrium,  no convective
overshooting, and 2\,\kms\ microturbulence. Starting from these initial estimates, 
we fixed the effective temperature ($T_{\rm{eff}}$) by imposing the excitation equilibrium
for \ion{Fe}{i} lines. For the microturbulence ($\xi$), we required no trend of the abundances versus
equivalent width (EW) for \ion{Fe}{i} lines. Finally, for the surface gravity
($\log~g$), we required that ionization equilibrium is achieved between
the abundances derived from the \ion{Fe}{i} \citep[adopting a non-LTE correction 
for \ion{Fe}{i} of 0.1\,dex from;][]{lindetal12} and \ion{Fe}{ii} lines.
A final iteration of the procedure was made with a model atmosphere computed
with these new parameters. The stellar parameters of the two program stars 
obtained by these procedures are listed in Table~\ref{tab3}.

\begin{table}[htbp]
  \caption{Stellar parameters of the targets.}
  \label{tab3}
  \centering
  \tiny
  \begin{tabular}{c c c c c} 
\hline
\noalign{\smallskip}
Object &$T_{\mathrm{eff}}$ $\mathrm{[K]}$ & $\log g$ $\mathrm{[cgs]}$ & $\xi$ $\mathrm{[km/s]}$&$\mathrm{[Fe/H]}$\\
\noalign{\smallskip}
\hline
\noalign{\smallskip}
CS~22186-005& $6160$ &$2.60$ &$3.40$ &$-2.70$\\
CS~30344-033& $6100$ &$2.70$ &$3.20$ &$-2.90$\\
\hline
\noalign{\smallskip}
  \end{tabular}
  \end{table}

The choice of using a temperature based on the excitation equilibrium
(instead of inferred from photometry) ensures that even if the stars
were RR-Lyrae stars, for which we have no present indication that they
are, the adopted temperature would be relevant for the pulsational phase
at which the star was observed \citep[see, e.g.,][]{Hansen11}.

\subsection{Line list and equivalent width measurement}

We compiled the line lists from \citet{aokietal07}, \citet{hayeketal09},
\citet{roedereretal10}, \citet{sbordoneetal10}, and \citet{yongetal13} for the measurements of
the atomic absorption lines in the stellar spectra. The $\log~gf$ values
for the \ion{Fe}{ii} and \ion{Ni}{i} lines were taken from
\citet{melendezbarbuy09} and \citet{woodetal14}, respectively.

\begin{figure}[htbp]
   \centering
   \includegraphics[bb=-30 410 483 721,width=11cm,clip]{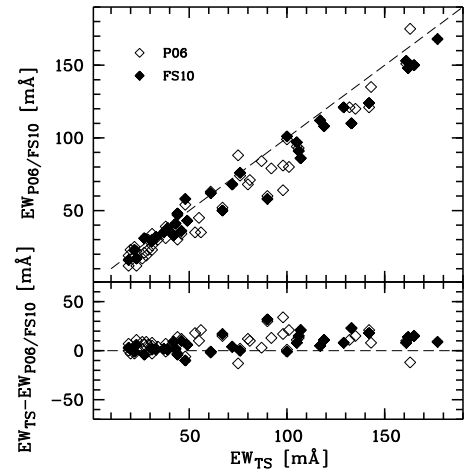}
\caption{Comparison of the measured EWs in this study (TS) and in the studies of \citet{prestonetal06} and \citet{forsneden10}.}
         \label{eqwcomp}
   \end{figure}

Equivalent width measurements were accomplished by fitting a Gaussian
profile to the lines simultaneously with a straight-line continuum,
where the continuum and line regions were chosen interactively. 
We compared our EW measurements with those of \citet{prestonetal06} and 
\citet{forsneden10} for CS~22186-005, as seen in Fig.~\ref{eqwcomp}.
The EWs measured in this study agree with the values obtained in the previous studies.
From the EW measurements, abundances were derived using the code WIDTH9 
\citep{sbordoneetal04, kurucz05}, which uses ATLAS9 model atmospheres to
compute line formation in LTE. 

\subsection{Spectrum synthesis}

Line lists that include hyperfine structure (HFS) splitting for Sc, Mn, Co,
and Ba were downloaded from Robert~Kurucz's web
page\footnote{http://kurucz.harvard.edu/linelists.html}. Since the EW
measurement method is not appropriate for lines where HFS splitting has
to be taken into account, we applied the spectrum synthesis method for
the Sc, Mn, Co, and Ba lines, producing synthetic spectra with the code SYNTHE 
\citep{kurucz93b, kurucz05}, and convolving the resulting spectra
with a Gaussian profile that includes the broadening effects caused by the
instrumental profile and the macroturbulence velocity. The abundances of
the species were then adjusted until the observed and synthetic spectra
agreed well.

\subsection{Uncertainties in abundances}

The random (statistical) uncertainties in the elemental abundances were taken to be
the standard deviation of individual line measurements ($\sigma$) over
the square root of the number of lines ($N$) for each element
($\sigma_{r}=\sigma/\sqrt{N}$). We adopted an uncertainty of $0.10$\,dex
for species that were measured with spectrum synthesis, or that had
only one detected line. This value takes into account the uncertainty in
the atomic data and the continuum placement. The random errors are listed in
Table~\ref{aburesu}.

\begin{table*}[htbp]
  \caption[]{Abundances of the species observed in CS~22186-005 and CS~30344-033.}
  \label{aburesu}
  \centering
  \tiny
  \begin{tabular}{l c r r c c r r r r c c}
    \hline
    \noalign{\smallskip}
            &                                     &\multicolumn{5}{c} {CS~22186-005} & \multicolumn{5}{c} {CS~30344-033}\\
   \noalign{\smallskip}  
   \hline
   \noalign{\smallskip}
   Species & $\log\epsilon_{\odot}$ $\mathrm{^a}$& N & $\log\epsilon$ &$\mathrm{[X/Fe]}$ & $\mathrm{[X/Fe]}$ $\mathrm{^b}$& $\sigma_{r}$&
 N &$\log\epsilon$ &$\mathrm{[X/Fe]}$& $\mathrm{[X/Fe]}$ $\mathrm{^b}$& $\sigma_{r}$ \\
    \noalign{\smallskip}
    \hline                         
    \noalign{\smallskip}           
    $\mathrm{[FeI/H]}$    &7.50&52      &4.80    &$-$2.70 &$\cdots$&0.02    &83      &   4.60 &$-$2.90 &$\cdots$&0.01    \\
    $\mathrm{[FeII/H]}$   &7.50&14      &4.95    &$-$2.55 &$\cdots$&0.03    & 6      &   4.69 &$-$2.81 &$\cdots$&0.04    \\
    Na~I                  &6.30&2       &3.57    &$-$0.03 &$-$0.33 &0.10    & 2      &   3.32 &$-$0.08 &$-$0.28 &0.08    \\
    Mg~I                  &7.55&5       &4.96    &   0.11 &   0.31 &0.12    & 6      &   4.92 &0.27    &   0.57 &0.05    \\         
    Al~I                  &6.46&2       &2.81    &$-$0.95 &$-$0.25 &0.01    & 2      &   2.80 &$-$0.76 &$-$0.06 &0.05    \\
    Si~I                  &7.54&1       &4.61    &$-$0.23 &$\cdots$&0.10    & 1      &   4.51 &$-$0.13 &$\cdots$&0.10    \\  
    Ca~I                  &6.34&5       &3.82    &   0.18 &$\cdots$&0.08    & 4      &   3.79 &0.35    &$\cdots$&0.03    \\                                         
    Sc~II$\mathrm{^{hfs}}$&3.07&3       &0.51    &   0.14 &$\cdots$&0.05    & 8      &   0.41 &0.15    &$\cdots$&0.07    \\                                     
    Ti~I                  &4.92&1       &2.98    &   0.76 &$\cdots$&0.10    & 1      &   2.62 &0.60    &$\cdots$&0.10    \\                                         
    Ti~II                 &4.92&18      &2.63    &   0.26 &$\cdots$&0.04    &28      &   2.72 &0.61    &$\cdots$&0.03    \\                                       
    V~II                  &4.00&$\cdots$&$\cdots$&$\cdots$&$\cdots$&$\cdots$& 2      &   1.72 &0.53    &$\cdots$&0.04    \\                                      
    Cr~I                  &5.65&5       &2.80    &$-$0.15 &$-$0.06 &0.04    & 6      &   2.67 &$-$0.08 &   0.01 &0.04    \\                                    
    Cr~II                 &5.65&$\cdots$&$\cdots$&$\cdots$&$\cdots$&$\cdots$& 1      &   2.57 &$-$0.27 &$-$0.22 &0.10    \\
    Mn~I$\mathrm{^{hfs}}$ &5.50&2       &2.38    &$-$0.42 &$\cdots$&0.04    & 3      &   2.02 &$-$0.58 &$\cdots$&0.03    \\
    Mn~II$\mathrm{^{hfs}}$&5.50&$\cdots$&$\cdots$&$\cdots$&$\cdots$&$\cdots$& 2      &   2.24 &$-$0.45 &$\cdots$&0.01    \\
    Co~I$\mathrm{^{hfs}}$ &4.91&$\cdots$&$\cdots$&$\cdots$&$\cdots$&$\cdots$& 4      &   2.35 &0.34    &$\cdots$&0.05    \\                                      
    Ni~I                  &6.22&22      &3.31    &$-$0.21 &$\cdots$&0.02    &12      &   3.37 &0.05    &$\cdots$&0.05    \\                                          
    Sr~II                 &2.91&2       &$-$0.78 &$-$1.14 &$-$0.94 &0.01    & 2      &$-$0.61 &$-$0.71 &$-$0.51 &0.02    \\ 
    Ba~II$\mathrm{^{hfs}}$&2.18&2       &$-$1.24 &$-$0.87 &$-$0.77 &0.02    &$\cdots$&$\cdots$&$\cdots$&$\cdots$&$\cdots$\\             
    \noalign{\smallskip}                                                                                 
    \hline                                                                                               
  \end{tabular}                                                                                          
  \tablefoot{\\                                                                                          
  \tablefoottext{a}{Solar abundance ratios are taken from \cite{lodders03}\\}                                                                                         
  \tablefoottext{b}{The non-LTE abundance ratios}}                                                                                  
  \end{table*} 

Systematic errors were estimated from the uncertainties in the stellar
parameters for each atomic species. The line-to-line scatter of the abundances, 
which is derived from \ion{Fe}{i} lines, is 0.12 for CS~22186-005 and 0.16 for CS~30344-033. 
We consider as the steepest acceptable slope in the abundance-excitation energy plane a value of $\pm 0.1/N$ per eV 
(where N is the number of lines). This translates into an uncertainty on $T_{\mathrm{eff}}$ of $\pm100$\,K. 
The abundances derived from the neutral and ionized species are still similar for an uncertainty 
of $\pm0.1$ in $\log~g$. The abundance-EW slope of the \ion{Fe}{i} lines plot is still lower than $0.01$ 
for an uncertainty of $\pm0.2$\,$\mathrm{km~s^{-1}}$ in $\xi$. For each model atmosphere, 
the stellar parameters $T_{\mathrm{eff}}$, $\log g$, and $\xi$ were varied
within an uncertainty of $\pm100$\,K, $\pm0.1$\,dex, and 
$\pm0.2$\, $\mathrm{km~s^{-1}}$. For example, 
to estimate the abundance uncertainties arising from the $T_{\mathrm{eff}}$, we
increased and decreased the value of $T_{\mathrm{eff}}$ by 100\,K from
its original value. We then computed new model atmospheres with these
values and re-computed the abundances for each species by fixing the other
atmospheric parameters. Then, we compared these abundances with the
abundances derived from the original stellar parameters. We applied the
same steps for $\log~g$ of $\pm0.1$\,dex and $\xi$ of $\pm0.2$\,
$\mathrm{km~s^{-1}}$. The systematic errors are listed in Table~\ref{temlogmic}. 

\begin{table*}[htbp]
   \caption[]{Abundance uncertainties due to uncertainties in the atmospheric parameters.}
   \label{temlogmic}  
   \centering
   \tiny
   \begin{tabular}{l r r r r r r r r r r r r}
      \hline
      \noalign{\smallskip}
              &\multicolumn{6}{c}{CS~22186-005}&\multicolumn{6}{c}{CS~30344-033}\\
      \hline
      \noalign{\smallskip}      
      Species
      &\multicolumn{2}{c}{$\Delta
        T_{\mathrm{eff}}$}&\multicolumn{2}{c}{$\Delta\log g$}&\multicolumn{2}{c}{$\Delta \xi$}&\multicolumn{2}
      {c}{$\Delta T_{\mathrm{eff}}$}&\multicolumn{2}{c}{$\Delta\log g$}&\multicolumn{2}{c}{$\Delta \xi$}\\
      \noalign{\smallskip}
      &\multicolumn{2}{c}{$\mathrm{[K]}$}&\multicolumn{2}{c}{$\mathrm{[dex]}$}&\multicolumn{2}{c}{$\mathrm{[km~s^{-1}]}$}&
      \multicolumn{2}{c}{$\mathrm{[K]}$}&\multicolumn{2}{c}{$\mathrm{[dex]}$}&\multicolumn{2}{c}{$\mathrm{[km~s^{-1}]}$}\\
      \noalign{\smallskip}
        &$+$100&$-$100&$+$0.1&$-$0.1&$+$0.2&$-$0.2&$+$100&$-$100&$+$0.1&$-$0.1&$+$0.2&$-$0.2\\
      \noalign{\smallskip}
      \hline     
      \noalign{\smallskip}                
Na~I &$-$0.07 &0.07    &   0.00 &0.00    &0.03    &$-$0.03 & $-$0.07 &0.07    &   0.00 &0.00    &0.03    &$-$0.02\\       
Mg~I &$-$0.06 &0.06    &   0.01 &0.00    &0.05    &$-$0.05 & $-$0.06 &0.07    &   0.01 &$-$0.01 &0.07    &$-$0.07\\       
Al~I &$-$0.08 &0.08    &   0.00 &0.00    &0.01    &$-$0.01 & $-$0.08 &0.08    &   0.01 &0.00    &0.02    &$-$0.01\\       
Si~I &$-$0.09 &0.08    &   0.00 &0.00    &0.04    &$-$0.04 & $-$0.09 &0.08    &   0.00 &0.00    &0.04    &$-$0.04\\       
Ca~I &$-$0.07 &0.06    &   0.01 &0.00    &0.03    &$-$0.02 & $-$0.07 &0.07    &   0.00 &0.00    &0.04    &$-$0.03\\       
Sc~II&$-$0.05 &0.06    &$-$0.05 &0.03    &0.01    &$-$0.01 & $-$0.08 &0.07    &$-$0.02 &0.03    &0.03    &$-$0.02\\       
Ti~I &$-$0.08 &0.09    &   0.00 &0.00    &0.00    &0.00    & $-$0.08 &0.09    &   0.00 &$-$0.01 &0.00    & 0.00  \\          
Ti~II&$-$0.05 &0.04    &$-$0.03 &0.04    &0.03    &$-$0.02 & $-$0.06 &0.06    &$-$0.03 &0.03    &0.05    &$-$0.05\\       
V~II &$\cdots$&$\cdots$&$\cdots$&$\cdots$&$\cdots$&$\cdots$& $-$0.07 &0.07    &$-$0.02 &0.02    &0.01    &$-$0.01\\           
Cr~I &$-$0.10 &0.09    &   0.01 &0.00    &0.01    &0.00    & $-$0.10 &0.10    &   0.01 &$-$0.01 &0.01    &$-$0.01\\       
Cr~II&$\cdots$&$\cdots$&$\cdots$&$\cdots$&$\cdots$&$\cdots$& $-$0.06 &0.06    &$-$0.02 &0.02    &0.05    &$-$0.04\\       
Mn~I &$-$0.11 &0.10    &0.01    &0.00    &0.01    &0.00    & $-$0.10 &0.11    &0.00    &$-$0.01 &0.00    &$-$0.01\\           
Mn~II&$\cdots$&$\cdots$&$\cdots$&$\cdots$&$\cdots$&$\cdots$& $-$0.06 &0.06    &   0.01 &$-$0.01 &0.03    &$-$0.02\\           
Fe~I &$-$0.09 &0.09    &   0.01 &0.00    &0.04    &$-$0.03 & $-$0.10 &0.10    &   0.01 &0.00    &0.05    &$-$0.04\\       
Fe~II&$-$0.02 &0.02    &$-$0.03 &   0.03 &0.02    &$-$0.02 & $-$0.01 &0.02    &$-$0.04 &   0.03 &0.02    &$-$0.02\\       
Co~I &$\cdots$&$\cdots$&$\cdots$&$\cdots$&$\cdots$&$\cdots$& $-$0.12 &0.11    &   0.01 &$-$0.01 &0.01    &   0.00\\           
Ni~I &$-$0.10 &0.10    &   0.02 &$-$0.02 &0.04    &$-$0.04 & $-$0.14 &0.14    &   0.02 &$-$0.02 &0.07    &$-$0.06\\       
Sr~II&$-$0.06 &0.06    &$-$0.03 &   0.03 &0.02    &$-$0.02 & $-$0.06 &0.06    &$-$0.03 &   0.03 &0.04    &$-$0.03\\       
Ba~II&$-$0.07 &0.07    &$-$0.03 &   0.02 &0.00    &$-$0.01 & $\cdots$&$\cdots$&$\cdots$&$\cdots$&$\cdots$&$\cdots$\\               
\noalign{\smallskip}                                                      
\hline                                            
\end{tabular}                                     
\end{table*}

\section{Evolutionary state and masses}

The derived effective temperatures and surface gravities of our two
program stars unambiguously classify them as HB stars. Their effective
temperatures place them close to the red edge of the RR-Lyrae
instability strip, whose position and dependence on metallicity is not
precisely known (see \citealt{prestonetal06}, \citealt{forsneden10}, 
and \citealt{Sandage10} for discussions on this point). This means that 
they {\it could} be pulsating RR-Lyrae stars. We inspected the spectra to see 
if we could detect any sign of P-Cyg or inverse P-Cyg profiles, which are 
present in the spectra of RR-Lyraes at some phases, but were unable to detect any. 
In addition, there is no published hint of photometric variability for either of the stars. 
Although absence of evidence is not evidence of absence, we 
assume that both stars are nonvariable red horizontal-branch
(RHB) stars. Our derived abundances are expected to be robust even if they are
RR-Lyrae stars, since the spectroscopically determined excitation
temperature we adopted would be appropriate for the observed phase.

\begin{figure}[htbp]
   \centering
   \includegraphics[clip=true]{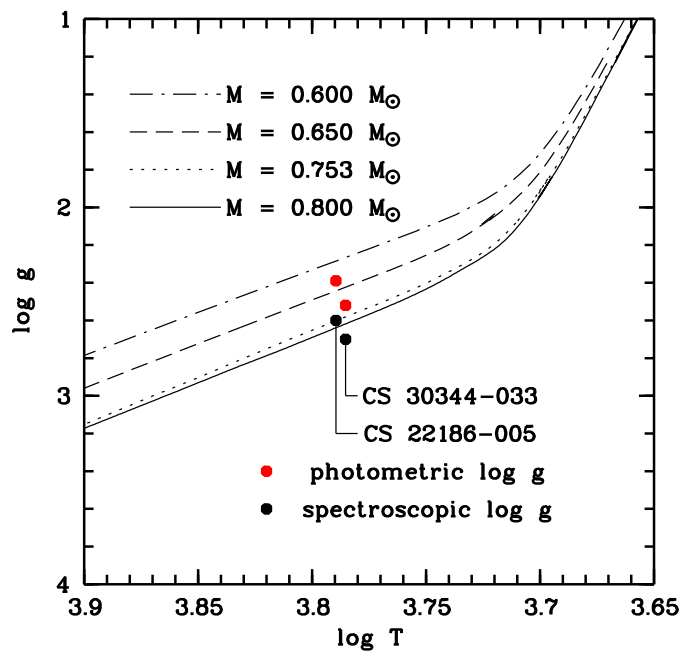}
\caption{Evolutionary tracks for HB stars of different masses and
metallicity $Z=0.00003$ in the $\log~T$ -- $\log~g$ plane \citep{cassisi}.
The black dots indicate the position of the two stars, with the spectroscopically derived $\log~g$, and the red
dots correspond to the "photometric" $\log~g$, derived from the
spectroscopic estimate, according to the empirical calibration of \citet{prestonetal06} }
         \label{hbtracks}
   \end{figure}

Fig.~\ref{hbtracks} shows evolutionary tracks for HB stars of
different mass and $Z=0.00003$ in the $\log~T$ -- $\log~g$ plane,
computed by \citet{cassisi}. It is clear from inspection that the tracks are
sufficiently well separated in this plane to allow determining the mass of each star. 
We used the same equation as employed by \citet{prestonetal06} to determine $\log~g$ from the
luminosity\footnote{$\log~g = \log (M/M_\odot)+4\log T_{\rm eff}-
\log (L/L_\odot)-10.607$}.

The two black dots in Fig.~\ref{hbtracks} identify the positions of
CS~22186-005 and CS~30344-033 in this plane, with our spectroscopically
derived parameters. In their study, \citet{prestonetal06} compared their
spectroscopic $\log~g$ for six RHB stars in the globular cluster M~15, with
the log g derived from the magnitudes of the stars, the cluster
distance, and an assumed median mass for the HB stars; they called this
the "photometric" $\log~g$. They derived a linear relation between the
photometric and the spectroscopic estimates of $\log~g$\footnote{$\log g_{\rm phot} =
\log g_{\rm spec} + 28.08 -7.655\log T_{\rm eff}$}. If we use this
relation to derive the log g for our program stars, the two stars move
toward lower gravities, shown in Fig.~\ref{hbtracks} as red dots.
Adopting the spectroscopic gravities, the two stars should have masses of
about $0.75~M_{\odot}$ and $0.8~M_{\odot}$, and about
$0.11~M_{\odot}$ lower masses for the "photometric" $\log~g$. We
refrain from precisely fitting the masses since our derived surface
gravity is uncertain by at least a factor of two. Still, this plot provides
a useful indication of the masses of the two stars.

\section{Abundance results} 
\subsection {The odd-Z elements}

We derived abundances for the odd-Z elements Na and Al for CS~22186-005
and CS~30344-033. Sodium abundances were determined for the two target stars
from EW measurements, using the $5889$\,{\AA} and $5895$\,{\AA} lines. 
\ion{Na}{i}~D lines are sensitive to non-LTE effects because of their strengths 
\citep{takedaetal03, andrievskyetal07}. We took into account non-LTE effects of 
$\sim-0.3$\,dex and $\sim-0.2$\,dex \citep{andrievskyetal07} for CS~22186-005 and 
CS~30344-033. We used the $3944$\,{\AA} and $3961$\,{\AA} lines to
determine Al abundances of the stars. The $3944$\,{\AA} line of \ion{Al}{i} is generally contaminated by the CH transition,
but this contamination is not observable in our spectra.
The abundances of aluminium were corrected for a large non-LTE offset of $\sim0.7$\,dex, 
as suggested by \citet{andrievskyetal08} because the aluminium resonance lines are affected 
by non-LTE effects at low metallicities. The derived light-odd element abundances are
listed in Table~\ref{aburesu}.

\subsection{The $\alpha-$elements}

The abundances of the $\alpha-$elements Mg, Si, Ca, and Ti were computed
for our two program stars. For CS~22186-005 the Mg abundances were determined from five \ion{Mg}{i} lines ($3838$\,{\AA}, $4702$\,{\AA},
$5172$\,{\AA}, $5183$\,{\AA}, and $5528$\,{\AA}) and for CS~30344-033 
from six lines ($3829$\,{\AA}, $3832$\,{\AA}, $3838$\,{\AA}, $5172$\,{\AA}, $5183$\,{\AA}, and
$5528$\,{\AA}). The Mg abundances were 
corrected using the non-LTE effects of $\sim0.3$\,dex calculated by \citet{andrievskyetal10}. 
The importance of deviations from LTE for Mg abundance 
was presented by \citet{andrievskyetal10} in the extended samples of metal-poor stars.

The Si abundances were measured from the $3905$\,{\AA} line in the spectra of the 
program stars. This line is sensitive to non-LTE effects, and the effect
increases with increasing effective temperature for metal-poor stars
\citep{shietal09}. The correction that results from the non-LTE effect
is $\sim0.05$\,dex. We did not take into account this non-LTE effect, because
the correction value is lower than the uncertainty on the Si
abundance estimates.

To determine Ca abundances, we used five \ion{Ca}{i} lines ($4226$\,{\AA}, $4302$\,{\AA},
$4318$\,{\AA}, $4454$\,{\AA}, $6439$\,{\AA})  for CS~22186-005
and four \ion{Ca}{i} lines ($4226$\,{\AA}, $4302$\,{\AA}, $4318$\,{\AA}, $4454$\,{\AA}) 
for CS~30344-033. The non-LTE effect on the Ca abundances of
the program stars was not taken into account \citep[$\sim0.05$\,dex, given by][]{spiteetal12}. 

We detected several \ion{Ti}{ii} lines and only one weak \ion{Ti}{i} line in the
observed spectra. The differences between the neutral and ionized titanium
are [\ion{Ti}{i}/\ion{Ti}{ii}]$=0.36\pm0.11$ for CS~22186-005, and
[\ion{Ti}{i}/\ion{Ti}{ii}]$=0.01\pm0.10$ for CS~30344-033.

\subsection{The iron-peak elements}

We have derived the abundances of the iron-peak elements Sc, V, Cr, Mn, Co,
and Ni for CS~30344-033. The abundances of Sc, Cr, Mn, and Ni were
determined for CS~22186-005. 

We used the \ion{Sc}{ii} lines $4247$\,{\AA}, $4314$\,{\AA}, and $4400$\,{\AA} 
to obtain the Sc abundance for CS~22186-005, and eight \ion{Sc}{ii} lines for 
CS~30344-033. The effects of HFS were taken into account for all of
these lines.

The vanadium abundance is determined from the $3517$\,{\AA}, and $3556$\,{\AA} 
lines of \ion{V}{ii} for CS~30344-033. We did not find any detectable vanadium 
line in our spectrum of CS~22186-005.

The Cr abundances were computed from five \ion{Cr}{i} lines for CS~22186-005,
and from six \ion{Cr}{i} lines and one \ion{Cr}{ii} line ($3409$\,{\AA}) for
CS~30344-033.  The difference between the neutral and ionized chromium is
[\ion{Cr}{i}/\ion{Cr}{ii}]=$0.10\pm0.11$ for CS~30344-033. At low metallicities, 
the overionization of \ion{Cr}{i} causes the deviations from LTE. The abundances of chromium 
were corrected for a non-LTE effect of $\sim0.3$\,dex, taken from \citet{bergemanncescutti10}.

The manganese abundance was measured from three \ion{Mn}{i} lines ($4030$\,{\AA}, 
$4033$\,{\AA}, $4034$\,{\AA}) and two \ion{Mn}{ii} lines ($3460$\,{\AA}, 
$3488$\,{\AA}) for CS~30344-033. For CS~22186-005, we determined the Mn abundance from
the $4030$\,{\AA} and $4034$\,{\AA} \ion{Mn}{i} lines. The effects of hyperfine splitting 
were taken into account for these lines. 

We used four Co lines ($3405$\,{\AA}, $3506$\,{\AA}, $3845$\,{\AA}, $3894$\,{\AA}) to 
measure the Co abundance in the spectrum of CS~30344-033. We accounted for the
effect of HFS for \ion{Co}{i} lines. We were unable to find any detectable Co line
in CS~22186-005.

No \ion{Ni}{i} line is confidently detectable in the HARPS spectrum of CS~22186-005.
We used the UVES spectrum that covers the UV region, where we measured twenty-two
\ion{Ni}{i} lines. Twelve \ion{Ni}{i} lines were used to obtain the Ni abundance from 
the spectrum of CS~30344-033.

\subsection{The neutron-capture elements}

The lines of the neutron-capture elements \ion{Sr}{ii} ($4077$\,{\AA} and
$4215$\,{\AA}) and \ion{Ba}{ii} ($4554$\,{\AA} and $4934$\,{\AA}) were detected
in the spectrum of CS~22186-005. We detected only two \ion{Sr}{ii} lines
($4077$\,{\AA} and $4215$\,{\AA}) in the spectrum of CS~30344-033. 
The non-LTE corrections are very important for Ba and Sr at low metallicities 
\citep{andrievskyetal09, andrievskyetal11}. To correct the Sr and Ba abundances, 
we accounted for non-LTE effects of $\sim 0.2$\,dex and $\sim 0.1$\,dex from 
\citet{andrievskyetal11} and \citet{andrievskyetal09}, respectively. In addition, 
the effect of HFS was taken into account for Ba~lines, assuming a solar isotopic ratio.                                                                                  
                                                                                                                                                                                                                 
\section{Discussion}                                                                     
                                                                                                         
Using high-resolution spectroscopy, we have derived elemental abundance
ratios for two very metal-poor HB stars, both of which have derived
metallicities lower than $\mathrm{[Fe/H]}=-2.5$. 

Our derived atmospheric parameters of CS~22186-005 agree well 
with those reported in the previous studies of \citet{prestonetal06} and \citet{forsneden10}. 
The differences between our measurements and the $T_{\mathrm{eff}}$, $\log g$, $\xi$, 
and [\ion{Fe}{i}/H] of \citet{forsneden10} are 40\,K, $-0.15$\,dex, $-$0.2\,\kms, and $-0.05$\,dex, 
respectively. Our $T_{\mathrm{eff}}$ is lower than the value reported by \citet{prestonetal06} by 90\,K. 
We derived a $\xi$ that is higher by 0.3\,\kms than that given by \citet{prestonetal06}.
The $\log~g$ and [\ion{Fe}{i}/H] in this study are the same as those of \citet{prestonetal06}.

\begin{figure}[htbp]
   \centering
   \includegraphics[bb=37 450 543 751,width=9cm,clip]{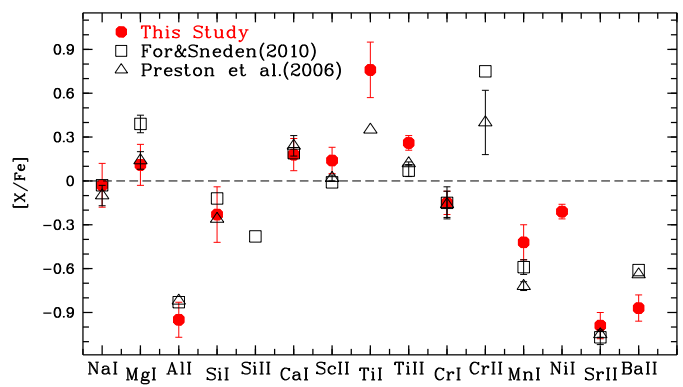}
   \caption{Comparison for CS~22186-005 of the abundance ratios derived from this analysis 
   with the previous studies of \citet{prestonetal06} and \citet{forsneden10}. The error bars show 
   the random uncertainties in the abundance ratios.}
   \label{cs22186}
   \end{figure}

The abundances of CS~22186-005 generally agree well with the study of \citet{forsneden10}, except for Mg and Ti.
For the magnesium abundance, \citet{forsneden10} used the two of Mg~I~b lines: $5172$\,{\AA} and $5183$\,{\AA}.
We have five Mg lines, which include the two \ion{Mg}{i}~b lines. If we use only the two Mg lines, our Mg abundance
agrees well with the value reported by \citet{forsneden10}. This shows that, as expected, the different Mg abundances 
are due to the lines used during the analysis. Our Ti abundance of CS~22186-005, 
derived from 18 \ion{Ti}{ii} lines, agrees reasonably well with the metal-poor halo stars, 
although it is somewhat higher than those of the previous studies. We show a comparison the abundance 
ratios of CS~22186-005 with those of the previous studies in Fig.~\ref{cs22186}.

In general, the abundance ratios that we report
here follow the Galactic trends traced by RGB and TO stars
\citep{yongetal13, roedereretal14b}, confirming previous claims that
HB stars are indeed reliable tracers of Galactic chemical evolution.

\subsection{The odd-Z elements}
\begin{figure*}[htbp]
   \centering
   \includegraphics[bb=29 508 615 735,width=15.5cm,clip]{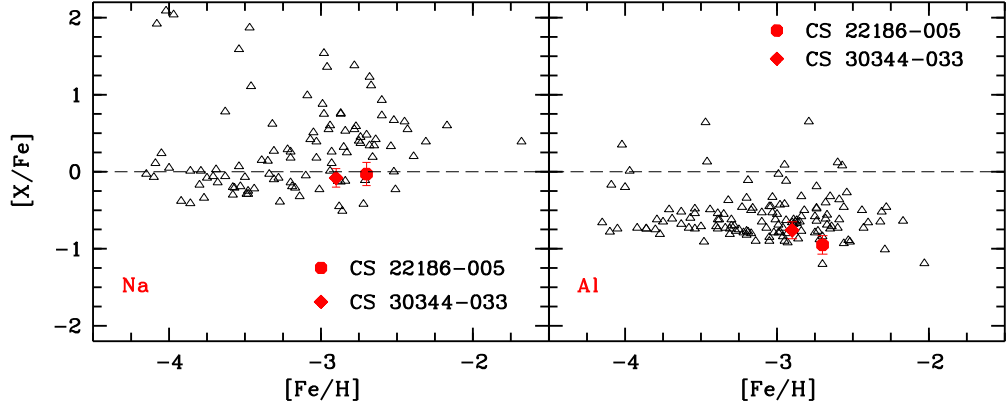}
      \caption{Light-odd elements (Na and Al) abundance ratios as a function of $\mathrm{[Fe/H]}$. 
Open triangles show the Galactic giant and turnoff stars \citep{yongetal13}. The error bars
show the random errors for each program star.}
         \label{lightoddelements}                                                                         
   \end{figure*}

The abundances of the odd-Z elements Na and Al decrease with decreasing
metallicity. The amount of Na and Al depends on the neutron excess, which
is produced in the CNO cyle during He burning \citep{kobayashietal06}.
We show $\mathrm{[(Na, Al)/Fe]}$ vs $\mathrm{[Fe/H]}$ in
Figure~\ref{lightoddelements}. The Na and Al abundance ratios of our
target stars are compatible with the trend of the other metal-poor
stars \citep{yongetal13}. 

\subsection{The $\alpha-$elements}
\begin{figure*}[htbp]
   \centering
   \includegraphics[bb=29 335 615 735,width=15.5cm, clip]{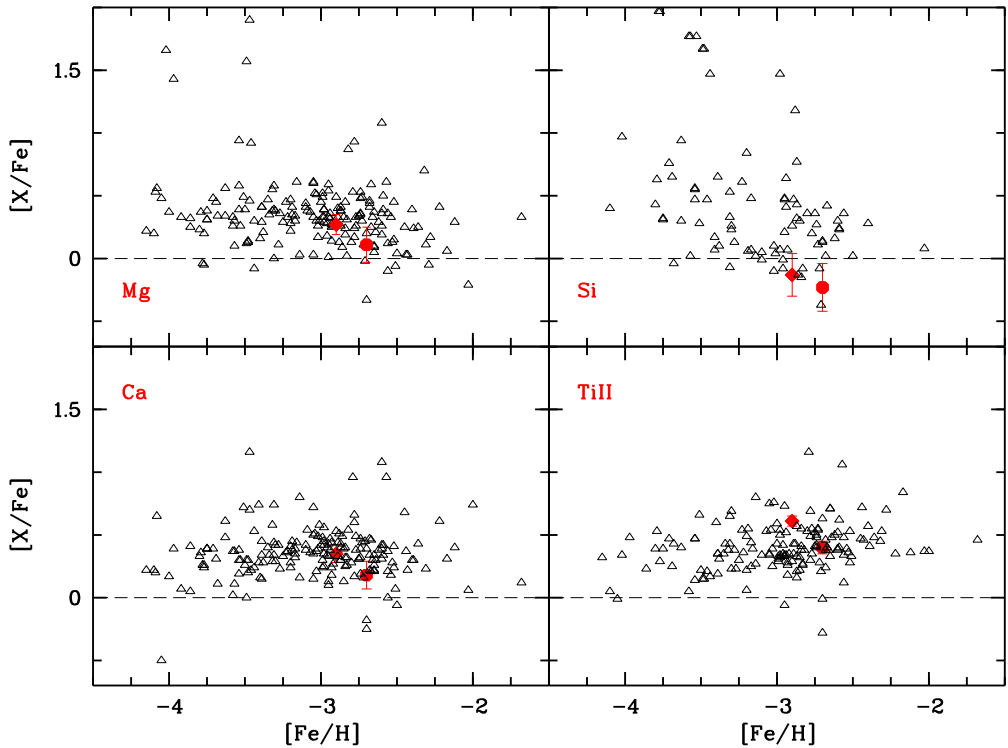}
      \caption{$\alpha-$element (Mg, Si, Ca, Ti) abundance ratios
of the program stars, compared with the Galactic trend. 
Symbols and references are the same as in Figure~\ref{lightoddelements}.}
         \label{alphaelements}
   \end{figure*}

Generally, metal-poor stars are enriched in $\alpha-$elements, because of
the lack of the contribution of Type~Ia at these metallicities. As
expected, the Mg, Ca, and Ti abundance ratios to iron of our target stars are
overabundant with respect to the solar values. However, the Si abundance
ratios of both target stars are underabundant with respect to the solar
values. For CS~22816-005, our measurement confirms those of
\citet{prestonetal06} and \citet{forsneden10}. These abundance values
are lower than the general trend, as shown in
Figure~\ref{alphaelements}. We stress that the Si abundance is
based on a single line, the $3906$\,{\AA} line, as is the case in other
very metal-poor RHB and TO stars.

\subsection{The iron-peak elements}
Sc abundance for the metal-poor stars follow a slightly decreasing trend with metallicity, 
as shown in Fig.~\ref{ironpeakelements}. The abundance results of Sc for our samples are  
compatible with the observed trend of this element. The V abundance, derived from single 
ionized vanadium, of CS~30344-033 is $0.53\pm0.04$. These overabundances 
in the vanadium abundance are unreliable because of the unresolved blends in the UV spectral 
range \citep{barklemetal05}. The metal-poor stars exhibit underabundances of Cr and Mn with 
decreasing metallicity \citep{cayreletal04, laietal08}. \citet{roedereretal14b} presented the 
trend of Cr and Mn abundances for subgiant, red giant, main-sequence, and HB stars. The Mn 
abundances of HB stars decrease with decreasing temperature. The chromium abundances are 
close to zero for the higher temperatures, while they decrease with decreasing temperature. 
Our HB samples almost agree with this trend of Cr and Mn. The $\mathrm{[Co/Fe]}$ displays 
an increase with decreasing metallicity \citep{cayreletal04}. The cobalt abundance of $0.34\pm0.05$ 
agrees with the observed trend, as seen in Fig.~\ref{ironpeakelements}.

\subsubsection{Evidence for scatter in $\mathrm{[Ni/Fe]}$?}
\begin{figure*}[htbp]
   \centering
   \includegraphics[bb=29 165 615 735,width=15.5cm, clip]{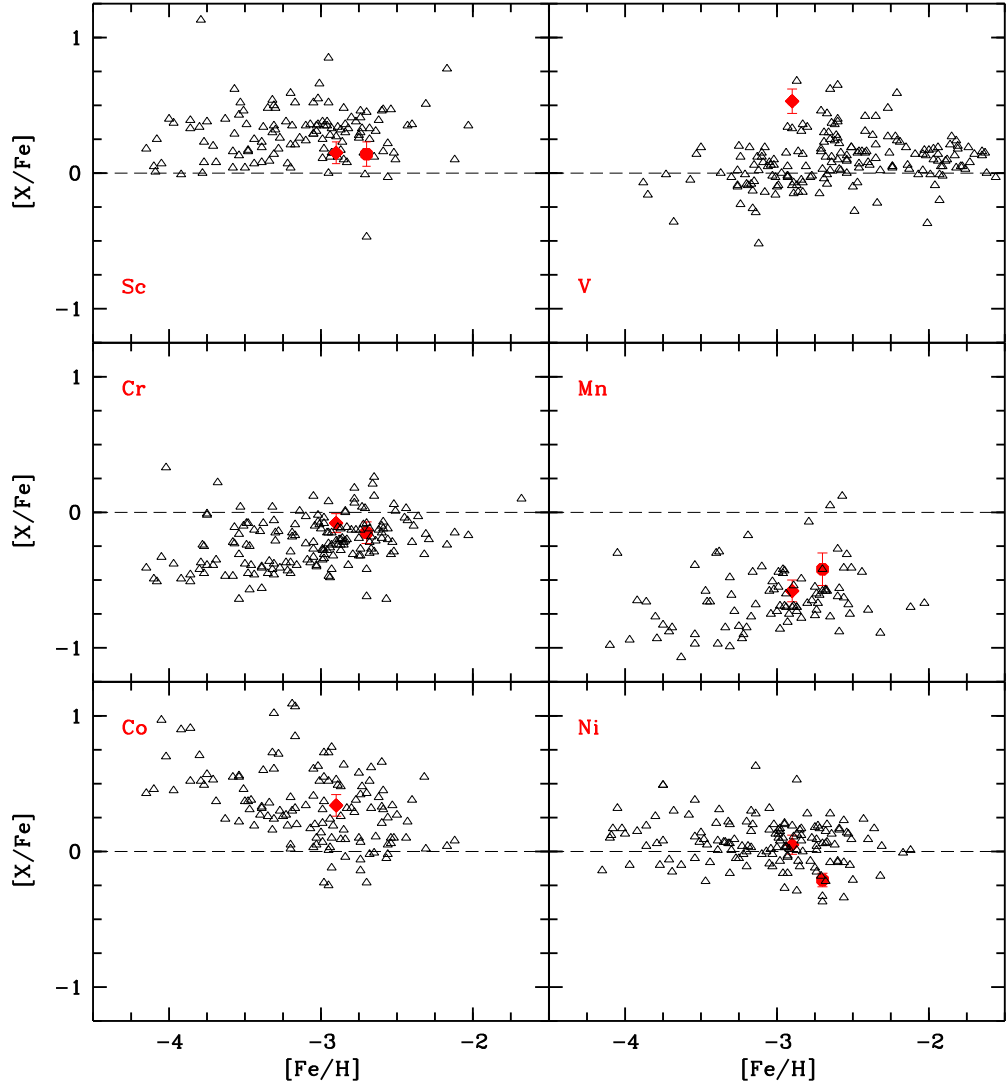}
      \caption{$\mathrm{[Sc, Cr, Mn, Co, Ni/Fe]}$ vs $\mathrm{[Fe/H]}$. Symbols and references are the same as in Figure~\ref{lightoddelements}.}
         \label{ironpeakelements}
   \end{figure*}

The $\mathrm{[Ni/Fe]}$ for CS~22186-005 was determined here for the first time. 
The value $\mathrm{[Ni/Fe]}=-0.21\pm 0.02$ was determined from 22 lines in the UV range. 
From Fig.~\ref{ironpeakelements} it can be appreciated that although the general trend with 
$\mathrm{[Fe/H]}$ is constant, $\mathrm{[Ni/Fe]}=0.0$, as expected for elements with the same 
nucleosynthetic origin, the scatter around this value is quite large. 
If we only consider the Ni abundances in the 38 stars analyzed by \citet{yongetal13},
neglecting the data re-analyzed from the literature, the scatter in $\mathrm{[Ni/Fe]}$ is $0.2$\,dex much larger
than what is expected from the errors in the individual measurements.
If we consider the whole sample of 190 stars of \citet{yongetal13},
which we took as comparison sample, the scatter is 0.17\,dex.
This scatter is higher than that in other samples of metal-poor stars.
For example, in the sample of \citet{bonifacio09} it is only of 0.06\,dex, 
and in the large sample of \citet{roedereretal14b} it is only 0.10\,dex.
This raises the question if the scatter in $\mathrm{[Ni/Fe]}$ is real. Our
measurement of Ni in CS~22186-005 is almost a factor of two below the mean of the rest of the sample.

In the sample of HB stars of \citet{forsneden10} there are three BHB stars (HD~60778, HD~74721, HD~93329) 
for which Ni could be measured only from \ion{Ni}{ii} lines, and they imply $\mathrm{[Ni/Fe]}$ between --0.4 and --0.3. 
Other metal-poor stars that have $\mathrm{[Ni/Fe]}\le -0.2$ are in the sample of \citet{roedereretal14b} 
CS~22174-020 and CS~22964-183, in the sample of \citet{yongetal13} we find CS~22953-037, CS~31061-032,
CS~22169-035, HE~1012-1540, HD~186478, HE~2232-0603, CS~29499-060, HE~1207-3108, and HE~2142-5656.

Among HB stars are a few stars in the sample of \citet{forsneden10} with $\mathrm{[Ni/Fe]}$ 
considerably higher than the mean (CS~ 22878-121, CS~22940-70, CS~22948-006, CS~22944-039, CS~22951-077, and 
CS~22886-043). The mean $\mathrm{[Ni/Fe]}$ of these stars is $+0.53$, with
an r.m.s scatter of 0.13\,dex. We point out that there are three stars
with high $\mathrm{[Ni/Fe]}$ in the sample of TO stars of \citet{Bonifacio12}, in
particular SDSS~J082521+040334, with $\mathrm{[Ni/Fe]}=+0.48$, measured from
four \ion{Ni}{i} lines, but also SDSS~J233113-010933, SDSS~J230814-085526,
SDSS~J154246+054426, and SDSS~J113528+010848, which span the range $+$0.21
to $+$0.32 \relax in $\mathrm{[Ni/Fe]}$, all from several \ion{Ni}{i} lines.
Furthermore, \citet{cohenetal13} reported that the Ni abundance ratios
of HE~1422-0818, HE~1434-1442, HE~0411-5725, and HE~1320+0139 have a
range of $+$0.42 to $+$0.76. Two stars (CS~22943-137 and G004-036) with
high $\mathrm{[Ni/Fe]}$ were discovered by \citet{roedereretal14b}. The nickel
abundance ratio of G004-036 was derived from eight lines as $\mathrm{[Ni/Fe]}=+0.57\pm0.11$. 
Finally, in the full sample of \citet{yongetal13} nine stars have $\mathrm{[Ni/Fe]}\ge+0.3$:
CS~30339-069, CS~29518-043, HE 1424-0241, CS~22948-093,CD $-24^\circ$~17504,
HE~0049-3948, HE~0228-4047, HE~1506-0113, and HE~2247-7400.

Is this span of a factor of three in $\mathrm{[Ni/Fe]}$ real?
The data ought to be analyzed in more detail before a firm conclusion on this question can be reached.
It is perhaps sobering to note that the most Ni-rich
star in the sample of \citet{yongetal13}, CS~29518-043, with $\mathrm{[Ni/Fe]}=+0.63$,
is a reanalysis of the equivalent widths of \citet{bonifacio09}, who derived 
$\mathrm{[Ni/Fe]}=+0.07$. In addition, for CS~30339-069 and CS~22948-093, 
\citet{yongetal13} derived $\mathrm{[Ni/Fe]}$ ratios much higher
than those derived by \citet{bonifacio09}
from the same equivalent widths.
On the side of low $\mathrm{[Ni/Fe]}$ stars in the sample
of \citet{yongetal13} for  CS~29499-060, they provide 
$\mathrm{[Ni/Fe]}=-0.21$, while \citet{bonifacio09} provide $+0.19$, 
for CS~22953-037 \citet{yongetal13} give $\mathrm{[Ni/Fe]}=-0.37$, 
while \citet{bonifacio09} give $+0.04$ and for CS~31061-032 
 \citet{yongetal13} give $-0.34$, while \citet{bonifacio09} give $+0.03$.
A homogeneous reanalysis of the available data
is necessary before a firm conclusion on the reality of the scatter in $\mathrm{[Ni/Fe]}$ can be drawn.
 
\subsection{The neutron-capture elements}

\begin{figure*}[htbp]
   \centering
   \includegraphics[bb=29 508 615 735,width=15.5cm,clip]{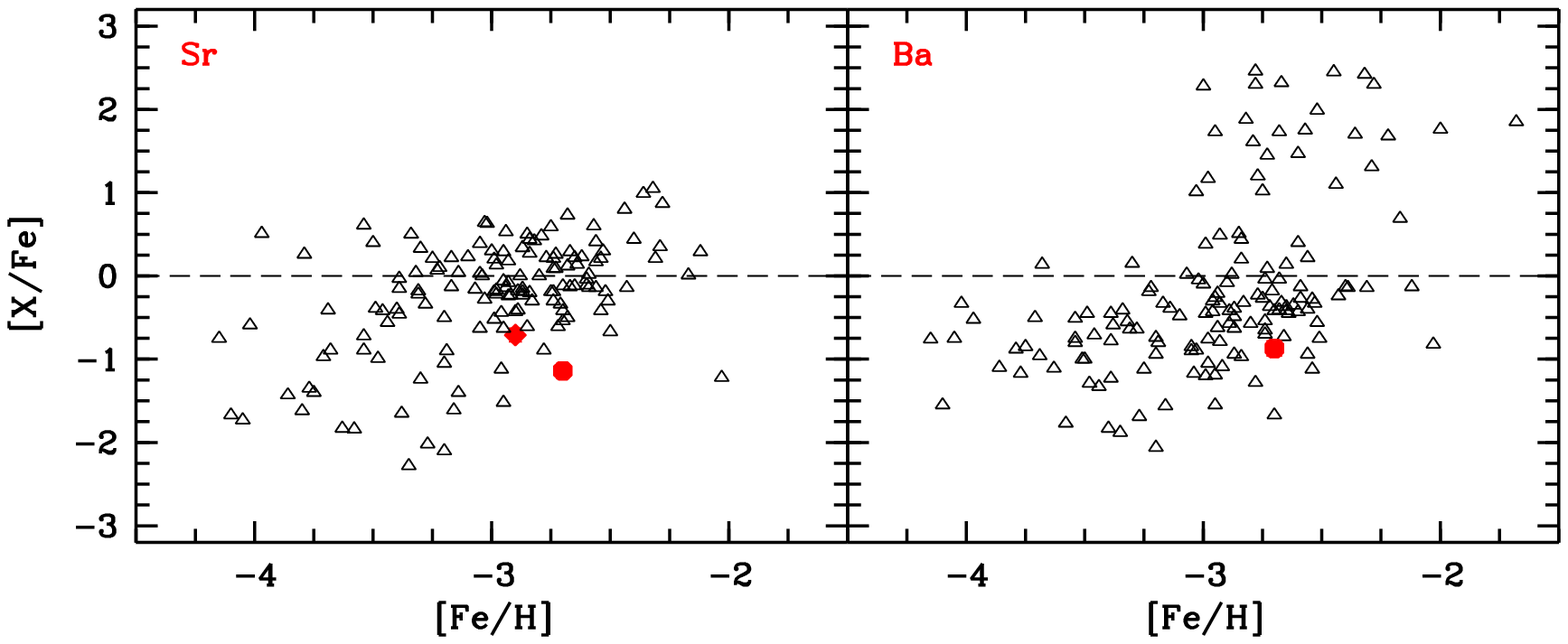}
      \caption{$\mathrm{[Ba, Sr/Fe]}$ vs $\mathrm{[Fe/H]}$. Symbols and references are the same as in Figure~\ref{lightoddelements}.}
         \label{neutroncaptureelements}
   \end{figure*}

We determined the abundances of the light and heavy neutron-capture 
elements Sr and Ba for CS~22186-005, while we only determined the Sr 
abundance for CS~30344-033. In both our HB stars the ratios of Sr 
and Ba to iron are lower than solar, which is not unusual among metal-poor stars. 
In Figure~\ref{neutroncaptureelements}, we compare the $\mathrm{[Sr,Ba/Fe]}$ 
abundance ratios as a function of $\mathrm{[Fe/H]}$ for our stars 
with those of the other metal-poor stars. The barium abundance of CS~22186-005 and 
strontium abundance of CS~30344-033 follow the general trend that decreases with 
decreasing metallicity, although both are on the low side.

\begin{figure*}[htbp]
   \centering
   \includegraphics[bb=37 450 543 751,width=9cm,clip]{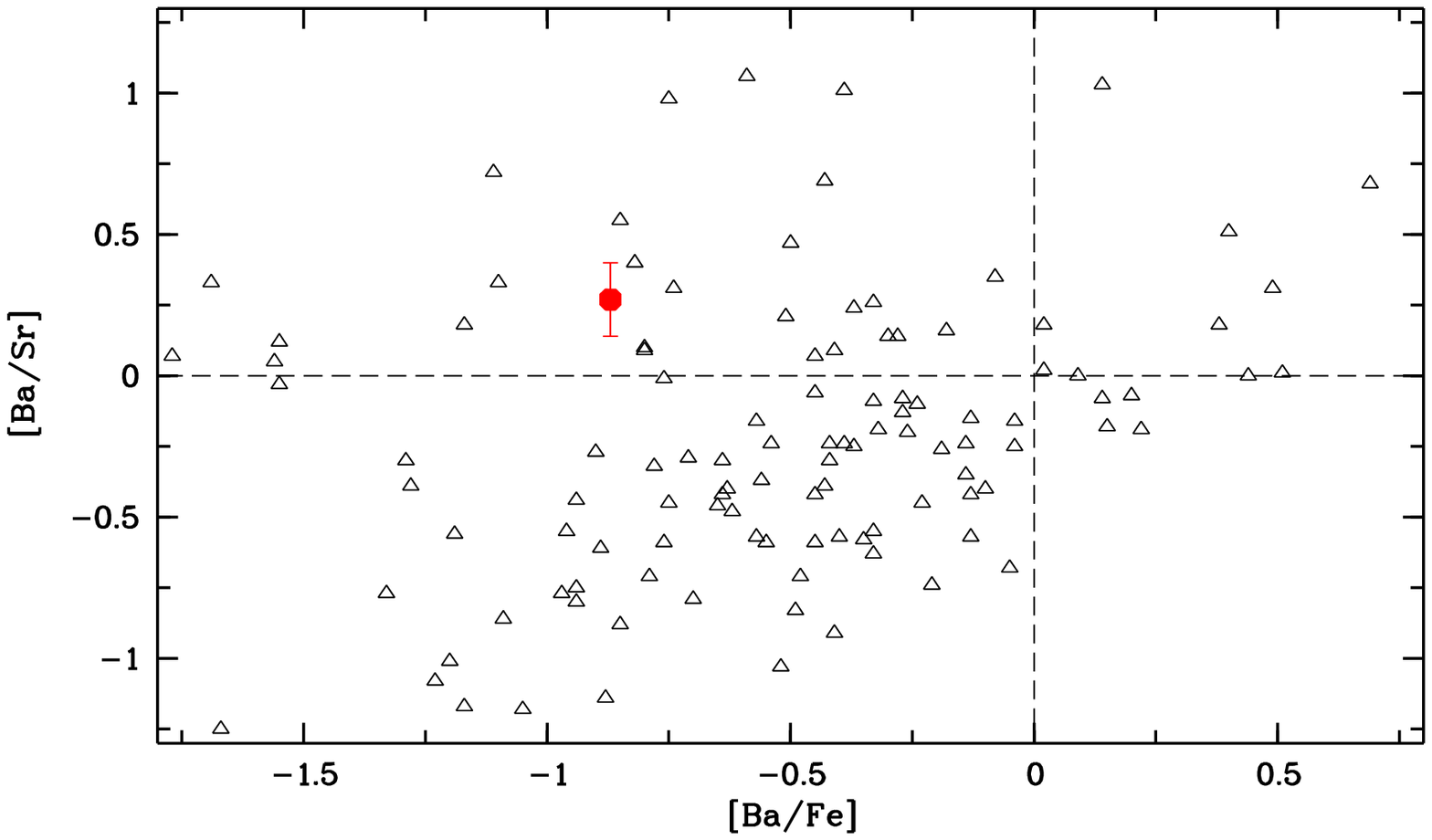}
      \caption{$\mathrm{[Ba/Sr]}$ vs $\mathrm{[Ba/Fe]}$. Symbols and references are the same as in Figure~\ref{lightoddelements}.}
         \label{BaSr}
   \end{figure*}

Based on the pioneering observations of \citet{spite78}, \citet{truran81} suggested 
that in low-metallicity stars all the neutron-capture elements are 
formed by the $r-$process\footnote{rapid neutron-capture, i.e., the time between 
successive captures of neutrons is shorter than with respect to the time for $\beta$ decay.}.
However, two decades later, it became clear that this view might be too simplistic.
\citet{travaglioetal04} concluded that although the classic 
$s-$process\footnote{slow neutron-capture, i.e., the time between
neutron-captures is longer than the time for $\beta$ decays. The classic
$s-$process comprises the main $s-$ process, responsible for the bulk
of the light neutron-capture elements in the solar system and
the weak $s-$process, characterized by a high neutron-to-seed nuclei
ratios, this process takes place in metal-poor stars.} could not account for the observations of
Sr, Y, Zr and their ratios to  heavy neutron-capture elements (Ba, Eu), neither was the $r-$process acting alone. 
They thus suggested that  Sr, Y, and Zr are formed via $s-$process in 
low-metallicity massive stars and called this process lighter element primary process (LEPP).
\citet{francoisetal07} noted that Sr, Y, and Zr exhibit an overabundance 
compared with Ba for metal-poor stars. \citet{snedenetal08} showed that the $\mathrm{[Ba/Sr]}$ abundance ratio increases 
as the $\mathrm{[Ba/Fe]}$ ratio increases. \citet{roedereretal14a} noted that two of their sample stars 
(CS~22891-200 and HE~1012-1540) had $\mathrm{[Ba/Sr]}>0$, and noted that the ratios are consistent with the 16 metal-poor 
stars of \citet{snedenetal08} that were enhanced by $r-$process nucleosynthesis. 
We show the $\mathrm{[Ba/Sr]}$ ratio (heavy/light) versus $\mathrm{[Ba/Fe]}$ 
for CS~22186-005 and the other metal-poor stars in Figure~\ref{BaSr}. 

As reported by \citet{forsneden10} and \citet{prestonetal06}, CS~22186-005 has a high $\mathrm{[Ba/Sr]}$ and 
low $\mathrm{[Ba/Fe]}$ ratio. Although this abundance pattern is exceptional among HB stars, among the
extremely metal-poor giants of \citet{francoisetal07} many exhibit similar ratios. The $\mathrm{[Ba/Sr]}$ ratios in 
CS~22186-005  was interpreted by \citet{forsneden10} as evidence for a multiplicity of neutron-capture synthesis  
processes.

\section{Conclusions}

Our analysis, based on original spectra of the newly discovered HB metal-poor star CS~30344-033 and of the
already known HB star CS~22186-005 confirmed that HB stars are reliable tracers of the Galactic
chemical evolution. Their higher luminosity, with respect to turnoff stars,
makes them suitable for probing the halo system of the Galaxy at larger distances.

The Ni abundance in CS~22186-005, determined here for the first time, implies
a $\mathrm{[Ni/Fe]}$ significantly below the mean value for most metal-poor stars. 
A closer inspection of the literature data suggests that taking the data
at face value, there might be room for an intrinsic scatter in the $\mathrm{[Ni/Fe]}$
ratios, but the large discrepancies in the $\mathrm{[Ni/Fe]}$ determinations
in the same stars by different authors implies that it is too early to draw a firm conclusion
on this question. 

Our radial velocity measurements for CS~22186-005, together
with the measurement of \citet{prestonetal06}, imply that
this star is a radial velocity variable. More observations are needed
to confirm this variability.

\begin{acknowledgements}
The authors thank an anonymous referee for the thoughtful comments.
We are grateful to Chris Sneden for discussion on their analysis of CS~22816-005. 
This work was supported by The Scientific and Technological Research Council of Turkey 
(T\"{U}B\.{I}TAK), the project number of 112T119. PB, and EC acknowledge support from 
the Programme National de Cosmologie et Galaxies (PNCG) of the Institut National de Sciences
de l'Univers of CNRS. EC is grateful to the FONDATION MERAC for funding her fellowsh. 
NC and LS acknowledge support from Sonderforschungsbereich 881 "The Milky Way
System" (subprojects A4 and A5) of the German Research Foundation (DFG). TCB acknowledges support 
from grant PHY 08-22648: Physics Frontiers Center/Joint Institute for Nuclear Astrophysics
(JINA), awarded by the U.S. National Science Foundation. LS acknowledges the support of Project 
IC120009 "Millennium Institute of Astrophysics (MAS)" of Iniciativa Científica Milenio del 
Ministerio de Economía, Fomento y Turismo de Chile.

\end{acknowledgements}

\bibliographystyle{aa}
\bibliography{scec}

\end{document}